\documentclass[reqno,12pt]{amsart}

\usepackage[centertags]{amsmath}
\usepackage{amsfonts}
\usepackage{amssymb}
\usepackage{amsthm}
\usepackage{newlfont}

\theoremstyle{plain}

\theoremstyle{definition}

\theoremstyle{remark}

\numberwithin{equation}{section}

\newcommand{\opunit}{\text{1}\kern-0.22em\text{l}}

\begin{document}

\begin{center}
\noindent{\large \bf  A quantum version of\\ free energy -
irreversible work relations} \\

\vspace{15pt}

{\bf Wojciech De Roeck}\footnote{Aspirant FWO, U.Antwerpen} and
{\bf Christian Maes}\footnote{email: {\tt
christian.maes@fys.kuleuven.ac.be}}\\
Instituut voor Theoretische Fysica\\ K.U.Leuven, Belgium.

\end{center}

\vspace{20pt} \footnotesize \noindent {\bf Abstract: } We give a
quantum version of the Jarzynski relation between the distribution
of work done over a certain time-interval on a system and the
difference of equilibrium free energies.  The main new ingredient
is the identification of work depending on the quantum history of
the system and the proper definition of various quantum ensembles
over which the averages should be made.  We also discuss a number
of different regimes that have been considered by other authors
and which are unified in the present set-up. In all cases, and
quantum or classical, it is a general relation between heat and
time-reversal that makes the Jarzynski relation so universally
valid.

\vspace{5pt}

 \vspace{20pt} \normalsize

\section{Introduction}
Thermodynamic potentials such as the Helmholtz free energy are
crucial in applications of thermodynamics. They give insight in
what processes are {\it a priori} workable, with what effects and
under what circumstances. Basically one is interested in two types
of information.  One is the expression of these potentials as a
function of system-parameters.  That determines the thermodynamic
landscape and it yields the thermodynamic forces. A second type of
information concerns the mutual relation between these potentials
and the link with available work and entropy-energy
transformations.  For example, for a system that can extract heat
from an environment at constant temperature $T$, the energy that
is available to do work is exactly the free energy $F \equiv V -
TS$, that is its energy $V$ minus the heat term $TS$ where $S$ is
the entropy of the system. Furthermore, to study its equilibrium
properties we should maximize the total entropy (of system and
reservoir) for a given energy contents but that again is
equivalent with minimizing the free energy at fixed temperature.

If, as often happens, no very reliable computation of the free
energy landscape can be made, the above provides immediate rescue.
It suffices to measure the work done under isothermal conditions
in changing the parameters of the system and it will be equal to
the free energy difference.  That however is only valid if the
thermodynamic process involved is quasi-static.  In other words,
the changes must be done very slowly, a situation that cannot be
hoped for in many cases.  It was therefore very useful that an
extended relation between free energy and work was proposed and
exploited in a series of papers since the pioneering work of
Jarzynski in 1997, \cite{ja1}.  That relation looks as follows:
\begin{equation}\label{jar}
e^{-\beta \Delta F} = \langle e^{-\beta W} \rangle
\end{equation}
In the left-hand side $\Delta F$ is what we want to know, the
difference in free energies between two equilibria with parameter
values $\kappa_f$ and $\kappa_i$. The right-hand side is an
average over all possible paths that take the system in
equilibrium for a certain parameter value $\kappa_i$ in its
initial Hamiltonian to a state where that parameter is changed
into $\kappa_f$. The work done $W$ depends on the path if the
process is not adiabatic (i.e. without heat transfer) or if it is
not quasi-static.  The protocol, i.e. the sequence of forcing in
the time-dependent Hamiltonian, is always kept fixed.

Derivations of the Jarzynski relation \eqref{jar} have been made
in various ways and in various approximations, see
\cite{Cro,ja1,ja2,ja3,ja4,ja5,MRV,MN}.   From such a relation free
energy differences can be measured even in situations where the
process of changing the parameters is not so well-controlled. That
has already been experimentally realized in e.g. molecular systems
\cite{Hum,Lip,Rit}.

A natural extension of \eqref{jar} concerns the quantum regime.
That appears possibly important and relevant when the system in
question should be treated with the methods of quantum mechanics.
In nano-devices the interplay between nanomechanics and
thermodynamics becomes all important.  Yet, a  more fundamental
reason to be interested in a quantum version of \eqref{jar} is the
question that it poses on the quantum nature of path-dependence.
We enter here the domain of quantum mechanics on histories.  It is
{\it a priori} not clear how to define quantum mechanical work
that depends on a path that the system has followed. In the
present paper we derive a quantum extension of \eqref{jar} where
we explicitly deal with that path-dependence and where we start
from a time-dependent unitary evolution on the level of the system
plus environment.

\section{Previous results}

Various proposals for quantum extensions of the Jarzynski relation
have appeared in the literature.  We briefly bring up some aspects
of such studies.

 In \cite{Tas} and following \cite{Kur} one introduces
the probabilities
\begin{equation}\label{probs}
p_{\alpha,\alpha'} \equiv \frac{e^{-\beta V_\alpha}}{Z(\beta)}
\,|\langle \varphi_{\alpha'}'|U|\varphi_\alpha\rangle|^2
\end{equation}
that the system is found in the $\alpha-$th eigenstate
$\varphi_\alpha$ of the Hamiltonian $H$ at an initial time (when
the system in in thermal equilibrium at inverse temperature
$\beta$) and then is found in the $\alpha'-$th eigenstate
$\varphi_{\alpha'}'$ of the Hamiltonian $H'$ at a later time.  The
operator $U$ in \eqref{probs} is the unitary operator for the time
inhomogeneous evolution during the whole period.  It is then an
easy computation, done in \cite{Tas}, that
\begin{equation}\label{tas}
\sum p_{\alpha,\alpha'} \exp[\beta V_{\alpha} - \beta
V_{\alpha'}'] =\frac{Z'(\beta)}{Z(\beta)}
\end{equation}
That resembles \eqref{jar} except for the important fact that the
left-hand side averages over an exponential of a total (in time)
energy difference.  In particular there is no concept here of a
path-dependent work.  One interpretation is that the system in
question is here really the total system (subsystem + heat bath)
and one should follow the change in energy over the whole system.
We think however that it is more useful to have a representation
of work in terms of the coordinates (and history) of the subsystem
only. After all, that is what free energies are all about
thermodynamically. A second interpretation is that one thinks of
the unitary evolution as working entirely on the subsystem itself
and the heat bath is completely absent except for the inverse
temperature $\beta$. That can be called the adiabatic regime and
we return to it in Section \ref{adi}.

 The
presentation in \cite{Muk} contains analogies both with what was
described in \cite{Cro} and with the adiabatic treatment of
\cite{Tas}.  In \cite{Cro} a derivation of \eqref{jar} was given
based on a time-inhomogeneous Markov process which satisfies, at
each time, the detailed balance relation for some energy function.
That can be lifted to the quantum regime when the Markov process
is seen as an effective description of a quantum system in contact
with a heat bath.  For example, the quantum weak coupling limit
exactly reproduces the {\it classical} Markov process as dynamics
for the system when in the energy basis for the system
Hamiltonian.  In that precise sense, \cite{Cro} was the first
quantum extension of the Jarzynski relation.  Here one deals with
an effective dynamics of the subsystem and we turn to in in
Section \ref{mama}.

Finally, in \cite{Yuk} the question of path-dependence of the work
is analyzed in an operator-setting and it is pointed out that
various ambiguities remain in the ordering of the operators. These
ambiguities only seem to disappear in a quasi-static limit which
unfortunately, is exactly the regime we are less interested in.
While it is in principle possible to define a work-operator for
the total system, the question in that setting remains whether its
projection on the subsystem remains useful and its spectrum
measurable.

In the present paper we deal with a unitary evolution over the
total system, subsystem plus reservoir, and we deal explicitly
with a path-dependent work.  That is new but the reason that the
Jarzynski relation \eqref{jar} is so universally valid remains the
same as for classical systems.  The basic observation is that the
entropy production can be identified with the source of
time-reversal breaking in the action governing the distribution of
system histories, see \cite{il,M,MN}.  We briefly state that point
here in a formal way to refer to it later when things become more
explicit.\\ We take the dynamics time-dependent through which work
$W$ is done on the system over a time-period while in contact with
a heat bath at constant inverse temperature $\beta$. The
time-dependent dynamics starts with an energy function $V_i$ and
at time $n$ the energy function is given by $V_f$.  Let
$q(\omega,\rho_i)$ denote the probability of a history
$(\omega_t)_0^n$ (on a certain level of description) of the system
started in the state $\rho_i$ at time zero. We choose $\rho_i =
\exp[-\beta V_i]/Z_i$.  We now reverse the protocol (the sequence
of forcing) and let $\tilde{q}(\omega,\rho_f))$ denote the
probability of the history $\omega$ when started in $\rho_f$ at
time zero.  We choose $\rho_f = \exp[-\beta V_f]/Z_f$ for the same
temperature but with a different energy function. For the
probability of the time-reversed trajectory $\Theta\omega \equiv
(\omega_{n-t})_0^n$, we then write
$\tilde{q}(\Theta\omega,\rho_f)$. Introducing the action
$\cal{L}$, we have
\begin{equation}\label{idea}
q(\omega,\rho) = \rho(\omega_0)e^{-\cal{L}(\omega)},\;
\;\frac{\tilde{q}(\Theta\omega,\rho_f)}{q(\omega,\rho_i)} =
e^{-R(\omega)}
\end{equation}
with
\[
 R(\omega) \equiv  \ln \rho_i(\omega_0) - \ln
\rho_f(\omega_n) + \tilde{\cal{L}}(\Theta\omega) - \cal{L}(\omega)
\]
Then,
\[
\ln \rho_0(\omega_0) - \ln \rho_n(\omega_n) = \beta[V_f(\omega_n)
- V_i(\omega_0)]  - \beta \Delta F,\quad \Delta F \equiv - \frac
1{\beta} \ln \frac{Z_f}{Z_i}
\]
The change in energy $V_f(\omega_n) - V_i(\omega_0)$ equals the
work $W$ minus the heat $Q$ that flows into the bath. On the other
hand, it can be argued that the source term of time-symmetry
breaking $\tilde{\cal{L}}(\Theta\omega) - \cal{L}(\omega)$ equals
the entropy production $\beta Q$, see \cite{MN}. Hence, we get
from \eqref{idea} that
\[
 \frac{\tilde{q}(\Theta\omega,\rho_n)}{q(\omega,\rho_0)} = e^{-\beta
 W(\omega) +\beta \Delta F}
 \]
But, by normalization, multiplying the above relation with
$q(\omega,\rho_i)$ and summing over all $\omega$ gives one; hence
\eqref{jar} obtains.\\

  What remains to be done is to give a
quantum expression for the above quantities and that is the
subject of the present paper.  In other words, we want to obtain
an algorithm, valid for a system subject to the laws of quantum
mechanics, through which we can measure the difference in
equilibrium free energies.  Moreover, we want this algorithm to be
formulated on the level of the subsystem.  That means that we must
trace out the heat bath from the equality \eqref{tas}.

\section{Formulation of the problem}\label{pa}

We model a possibly small quantum system in contact with a much
larger heat reservoir kept at fixed inverse temperature $\beta$.
The Hilbert space of the subsystem is denoted by $\cal{H}_S$ and
that of the environment by $\cal{H}_R$; both are assumed finite
dimensional. As usual the total real and self-adjoint Hamiltonian
is a sum of three contributions,
\begin{equation}\label{ham}
H_t = H^S_t + H^R + \gamma H^I
\end{equation}
 where the system-part $H_t^S$ is
parameterized by $t=0,1,\ldots,n$ and acts on $\cal{H}_S$. The
Hamiltonian of the reservoir $H^R$ (acting on $\cal{H}_R$) and the
coupling $H^I$ between system and reservoir are assumed fixed. We
will not need any explicit description of these terms.  Setting
$\gamma=0$ decouples system and reservoir.   The canonical density
matrices describing equilibrium for the decoupled system are
\begin{equation}\label{eq}
\rho^t \equiv \frac 1{Z_t} e^{-\beta H^S_t} \otimes \frac 1{Z^R}
e^{-\beta H^R}
\end{equation}
We are interested in the difference of Helmholtz free energies
\begin{equation}\label{helm}
\Delta F \equiv - \frac 1{\beta} \ln \frac{Z_n}{Z_0}
\end{equation}
 The
dynamics for the total system is unitary and time-dependent with
unitary operator
\[
U_t = e^{i\lambda H_t}
\]
acting on $\cal{H}_S \otimes \cal{H}_R$. The parameter $\lambda$
is real and sets the energy-time scale.\\

While the left-hand side of \eqref{jar} is clear and given by
\eqref{helm} the question is about the quantum version of the
right-hand side:  What is the averaging and what is the work?
Different reduced dynamics for the subsystem can be imagined that
are relevant in different types of regimes.

\section{Results}

\subsection{Effective regime}\label{mama}
Here we suppose that the dynamics for the subsystem  is described
via some effective dynamics.  There are various candidates but one
class of examples is obtained as the quantum analogue of a Markov
process on $\cal{H}_S$. These can be rigorously obtained under
various conditions and in various limiting regimes. Following
\cite{DS}, one can start with a time-dependent Hamiltonian $H_t$
and take the weak coupling limit. Obviously the driving protocol
has to vary on the same time scale as the dissipation processes
through contact with the reservoir. What results is a
time-inhomogeneous Markov process such that the instantaneous
generator at time $t$ satisfies detailed balance with respect to
$H(t)$.
 One way to implement that is to think of a
sequence $\varphi_{\alpha_O}^0\rightarrow
\varphi_{\alpha_1}^0\rightarrow
\varphi_{\alpha_1}^1\rightarrow\varphi_{\alpha_2}^1\rightarrow\ldots
\rightarrow
\varphi_{\alpha_{n-1}}^{n-1}\rightarrow\varphi_{\alpha_n}^{n-1}\rightarrow
\varphi_{\alpha_n}^{n}$ where alternating in time, the transition
 is either
thermal as for
$\varphi_{\alpha_{t}}^{t}\rightarrow\varphi_{\alpha_{t+1}}^{t}$,
and is modelled by a completely positive map $\Lambda_t$ which
satisfies the condition of detailed balance with respect to
$\rho_t \equiv e^{-\beta H^S_t}/Z_t$,
\begin{equation}\label{db}
\frac{\mbox{Tr }[P_{\alpha_{t}}^{t} \Lambda_{t}(
P_{\alpha_{t-1}}^{t})]} {\mbox{Tr }[P_{\alpha_{t-1}}^{t}
\Lambda_{t}( P_{\alpha_{t}}^{t})]} =
\exp[-\beta(V_{\alpha_{t}}^t-V_{\alpha_{t-1}}^t)]
\end{equation}
or is mechanical as for
$\varphi_{\alpha_{t}}^{t}\rightarrow\varphi_{\alpha_{t}}^{t+1}$.
That last transition is imagined instantaneously performed so that
we define the probability of a trajectory $\omega$ as the product
\begin{eqnarray}\label{pathprobD}
q_\beta^D(\omega) \equiv \mbox{Tr }[P_{\alpha_{n}}^{n}
\Lambda_{n}( P_{\alpha_{n-1}}^{n})] &&\mbox{Tr
}[P_{\alpha_{n-1}}^{n-1} \Lambda_{n-1}( P_{\alpha_{n-2}}^{n-1})]
\ldots\\\nonumber &&\ldots\mbox{Tr }[P_{\alpha_1}^{1} \Lambda_{1}(
P_{\alpha_{0}}^{1})]\frac{e^{-\beta V^0_{\alpha_0}}}{Z_0}
\end{eqnarray}
 Expectations
will be denoted by $\langle \cdot \rangle_D$. The total change in
energy is $\Delta V \equiv V_{\alpha_n}^n - V_{\alpha_0}^0 $ and
the total heat that flows in the heat bath in the thermal
transitions \eqref{db} is
\begin{equation}\label{hea}
Q(\omega) \equiv -\sum_{t=1}^{n}(V_{\alpha_{t}}^t
-V_{\alpha_{t-1}}^t)
\end{equation}
The total work is therefore defined as
\begin{equation}\label{wo}
W(\omega)\equiv Q(\omega) + \Delta V =
\sum_{t=0}^{n-1}(V_{\alpha_{t}}^{t+1}-V_{\alpha_{t}}^t)
\end{equation}
and is done over the transitions
$\varphi_{\alpha_{t}}^{t}\rightarrow\varphi_{\alpha_{t}}^{t+1}$.
We then have
\begin{equation}\label{corpocro}
\langle e^{-\beta W}\rangle_D = e^{-\beta \Delta F}
\end{equation}

The simplest way to prove \eqref{corpocro} is to use the relation
between entropy production and time-reversal as in
\cite{M,MN,il,Cro}.  Let
$\Theta\omega\equiv(\alpha_n,\ldots,\alpha_0)$ be the
time-reversed trajectory.  Similar to \eqref{pathprobD} we define
a path-space measure starting from $\rho_n$:
\begin{equation}\label{pathprobDt}
\tilde{q}_\beta^D(\Theta\omega) \equiv \mbox{Tr }[P_{\alpha_0}^{1}
\Lambda_{1}( P_{\alpha_{1}}^{1})] \ldots \mbox{Tr
}[P_{\alpha_{n-1}}^{n} \Lambda_{n}( P_{\alpha_{n}}^{n})]
\frac{e^{-\beta V^n_{\alpha_n}}}{Z_n}
\end{equation}
 and compute the ratio
\[
\frac{q_\beta^D(\omega)}{\tilde{q}_\beta^D(\Theta\omega)} =
e^{\beta (V^n_{\alpha_n} -
V^0_{\alpha_0})}\frac{Z_n}{Z_0}\frac{\mbox{Tr }[P_{\alpha_{n}}^{n}
\Lambda_{n}( P_{\alpha_{n-1}}^{n})]}{\mbox{Tr
}[P_{\alpha_{n-1}}^{n} \Lambda_{n}( P_{\alpha_{n}}^{n})]} \ldots
\frac{\mbox{Tr }[P_{\alpha_1}^{1} \Lambda_{1}(
P_{\alpha_{0}}^{1})]}{\mbox{Tr }[P_{\alpha_0}^{1} \Lambda_{1}(
P_{\alpha_{1}}^{1})]}
\]
By using detailed balance \eqref{db} at every time-step and the
definitions \eqref{hea}--\eqref{wo}, one arrives at
\[
\frac{q_\beta^D(\omega)}{\tilde{q}_\beta^D(\Theta\omega)}
=e^{\beta\Delta V - \beta \Delta F + \beta Q} = e^{\beta W - \beta
\Delta F}
\]
Apply to that relation the normalization condition
\[
\sum_\omega
q_\beta^D(\omega)\frac{\tilde{q}_\beta^D(\Theta\omega)}{q_\beta^D(\omega)}
= 1
\]
to conclude \eqref{corpocro}.  The proof above mimics exactly the
scenario of \eqref{idea}. The result is the very analogue of the
main identity by Crooks in \cite{Cro} but where the transition
rates in \eqref{pathprobD} have a quantum mechanical expression.

\subsection{Repeated measurements}\label{genc}

We come back to the set-up of \eqref{ham}--\eqref{eq}. Assume that
each $H^S_t$ is non-degenerate and has projections $P_\alpha^t$ on
its eigenstates
 $\varphi_\alpha^t$ with eigenstates $V_\alpha^t$.  A trajectory
or path for the subsystem is a sequence
$(\alpha_0,\ldots,\alpha_n)$ where each $\alpha_t$ runs over the
possible eigenstates of $H^S_t, t=0,\ldots,n$.  We now give a
probability measure on such trajectories which is obtained by
tracing out the quantum
mechanical probabilities for the whole system.\\
Let $P_E$ denote the projection on the energy-space in $\cal{H}_R$
for the reservoir Hamiltonian $H^R$ with energy $E$.  The
probability to find the total system initially in equilibrium for
\eqref{eq} and at later times in eigenstates
$\varphi_{\alpha_t}^t$ for the system and with energies $E_t$ for
the reservoir is given by
\begin{eqnarray}\label{pathprob}
&& p_\beta(\alpha_0,\ldots,\alpha_n;E_0,\ldots,E_n) \equiv
 \mbox{Tr }[G \, \rho^0 \, G^\star]\\\nonumber \mbox{ with } &&
G \equiv P_{\alpha_n}^n \otimes P_{E_n} U_{n} \ldots
P_{\alpha_1}^1 \otimes P_{E_1} U_1 P_{\alpha_0}^0 \otimes P_{E_0}
\end{eqnarray}
When viewed from the subsystem, the probability for trajectory
$\omega = (\alpha_0,\ldots,\alpha_n)$ is thus (let $e\equiv
(E_0,\ldots,E_n)$)
\begin{equation}\label{pathprobsys}
q_\beta(\omega) \equiv \sum_{e}
p_\beta(\alpha_0,\ldots,\alpha_n;e)
\end{equation}
and when conditioning on $\omega$, \eqref{pathprob} gives
expectations denoted as
\begin{equation}\label{expathc}
\langle g \rangle (\omega) \equiv \frac 1{q_\beta(\omega)}
\sum_{e} g(\omega,e) \;p_\beta(\omega;e)
\end{equation}
when $q_\beta(\omega)$ is non-zero.  Finally, the expectations in
the path-space measure \eqref{pathprobsys} are written as
\begin{equation}\label{expath}
\langle f \rangle \equiv \sum_{\omega} f(\omega)\; q_\beta(\omega)
\end{equation}

The change in energy for the subsystem corresponding to the path
$\omega$ is  $V_{\alpha_n}^n - V_{\alpha_0}^0$ where $V_\alpha^t$
is the energy of $\varphi_{\alpha}^t$. We define a path-dependent
work by the formula
\begin{equation}\label{work}
W(\omega) \equiv V_{\alpha_n}^n - V_{\alpha_0}^0 - \frac 1{\beta}
\ln \langle e^{-\beta (E_n-E_0)} \rangle (\omega)
\end{equation}

The interpretation follows the first law of thermodynamics.  To
change the parameters in the Hamiltonian $H^S_t$ isothermally some
heat must flow from the bath into the system.  That is the second
term in \eqref{work}.  We can  expect that the heat bath is
dispersionfree with respect to the subsystem in the sense that
through each step $\varphi_{\alpha_t}^t \rightarrow
\varphi_{\alpha_{t+1}}^{t+1}$ of the trajectory $\omega$, the
corresponding change in energies $E_{t+1}-E_t$ of the reservoir is
determined:
\begin{equation}\label{as}
 -\frac 1{\beta} \ln \langle e^{-\beta (E_n-E_0)} \rangle (\omega)
 \simeq
 \langle E_n-E_{0} \rangle (\omega)
 \end{equation}
 That gives the heat $Q$
 flowing into the reservoir.
 At the same time the energy in the subsystem changes, the first
term in \eqref{work}. Combined, \eqref{work} gives the work
performed on the subsystem.\\
 For  \eqref{pathprob} -- \eqref{work},
\begin{equation}\label{realJarz}
e^{-\beta \Delta F} = \langle \, e^{-\beta W} \,\rangle
\end{equation}
That means that the Jarzynski relation \eqref{jar} is unaffected
in the quantum regime when, in the averaging, the quantum
mechanical probabilities are used.  We will now verify
\eqref{realJarz}.\\

We apply again the ideas around
 \eqref{idea}.
We define the time-reversed path-space measure from
\eqref{pathprob} by reversing the order in which the
time-dependent dynamics is applied and by now starting from the
density matrix $\rho^n$ of \eqref{eq}:
\begin{eqnarray}\label{pathprobt}
&& \tilde{p}_\beta(\alpha_0,\ldots,\alpha_n;E_0,\ldots,E_n) \equiv
 \mbox{Tr }[\tilde{G} \, \rho^n \, \tilde{G}^\star]\\\nonumber \mbox{ with } &&
\tilde{G} \equiv P_{\alpha_n}^n \otimes P_{E_n} U_{1^\star} \ldots
P_{\alpha_1}^1 \otimes P_{E_1} U_n^\star P_{\alpha_0}^0 \otimes
P_{E_0}
\end{eqnarray}
It follows immediately that
\begin{equation}
\tilde{p}_{\beta}(\Theta\omega;\Theta e)=p_{\beta}(\omega;e)
e^{-\beta(V^{n}_{\alpha_n}-V^{0}_{\alpha_0}+E_n-E_0)}\frac{Z_{0}}{Z_{n}}
\end{equation}
and hence
\begin{eqnarray}
 \langle \, e^{-\beta W}
 \,\rangle
 &=& \sum_{\omega}q_\beta(\omega)e^{\beta(V^0_{\alpha_0}-V_{n}^{\alpha_n})}\langle
 e^{-\beta(E_n-E_0)}\rangle (\omega)\\\nonumber
 &=& \sum_{\omega,e} p_\beta(\omega,e) e^{\beta(V^0_{\alpha_0}-V^{n}_{\alpha_n})}
 e^{-\beta(E_n-E_0)}\\\nonumber
 &=& \frac{Z_{n}}{Z_{0}} \sum_{\omega,e} \tilde{p}_\beta(\Theta\omega,\Theta e)\\\nonumber
 &=& e^{-\beta \Delta F}
\end{eqnarray}
as required.\\

The repeated measurements introduce another aspect of randomness
in the distribution of work which is absent classically.  Unless
one is taking an effective dynamics like in Section \ref{mama},
one will always need to take care of that aspect to define in any
useful way what is meant by work that depends on the history of
the subsystem.

\subsection{Special cases}

There are a number of special
cases that we  treat separately.\\

\subsubsection{Adiabatic regime}\label{adi}
We consider only the subsystem that was initially
brought in thermal equilibrium at inverse temperature $\beta$ and
that from time zero on is isolated from the environment.  We take
thus the same set-up as in Section \ref{pa} except that we cut the
coupling with the reservoir. The initial density matrix is
\[
\rho_0 \equiv \frac 1{Z_0} e^{-\beta H^S_0}
\]
and the dynamics is unitary on $\cal{H}_S$ and here denoted by
$U^S_t$ with $t=0,1,\ldots,n$ changing as time proceeds. $U^S_t$
need not commute with $H^S_t$. Instead of \eqref{pathprobsys} we
now take the probability of trajectory $\omega
=(\alpha_0,\ldots,\alpha_n)$ to be
\begin{equation}\label{pathprobS}
q_\beta^S(\omega) \equiv
 \mbox{Tr }[P_{\alpha_n}^n  U_{n}^S \ldots
P_{\alpha_1}^1 U_1^S P_{\alpha_0}^0  \, \rho_0 \, (P_{\alpha_n}^n
U_{n}^S \ldots P_{\alpha_1}^1 U_1^S P_{\alpha_0}^0)^\star]
\end{equation}
with expectations $\langle \cdot \rangle_S$.  For $\omega$ the
change in energy of the subsystem is $V_n - V_0$ as was the first
term in \eqref{work}. Then

\begin{equation}\label{corpo}
\langle e^{\beta(V_{\alpha_0}^0 - V_{\alpha_n}^n)}\rangle_S =
e^{-\beta \Delta F}
\end{equation}

That identity is the generalization of Equation 2.7 in \cite{Tas}.
Note that \eqref{corpo} is true for an arbitrary family of unitary
operators defining the time-evolution. It can be obtained from the
following exact identity.
 Let $G$ be an operator on $\cal{H}_S$ and write Tr
$[G\,P^n_{\alpha}] \equiv G(\alpha)$. Then, as one easily checks,
\begin{equation}\label{forcorpo}
\sum_{\omega} G(\alpha_n)\, e^{\beta V_{\alpha_0}^0} \mbox{Tr
}[P_{\alpha_n}^n \Lambda_{n-1} (\ldots P_{\alpha_1}^1 (\Lambda_1
(P_{\alpha_0}^0  \, \rho_0 \,
P_{\alpha_0}^0))P_{\alpha_1}^1\ldots)] = \frac{\mbox{Tr }[G]}{Z}
\end{equation}
for all super-operators $\Lambda_t$ (acting linearly on density
matrices) that leave the identity invariant,
$\Lambda_t(1)=1$.\\
One can generalize \eqref{pathprobS} and \eqref{corpo} by choosing
here
\[
 \Lambda_t(A) = \sum_r
\mu_r^t\,\; U_r^S \;A\; {U_r^S}^\star
\]
with $\mu_r^t\geq 0$ and $\sum_r\mu_r^t = 1$, meaning that the
unitary $U_r^S$ is employed with probability $\mu_r^t$ at time
$t$. These $\Lambda$'s leave the identity invariant, so
\eqref{forcorpo} applies and
\begin{equation}
\sum_\omega \,q_\beta^e(\omega)\,e^{\beta(V_{\alpha_0}^0 -
V_{\alpha_n}^n)} = e^{-\beta \Delta F}
\end{equation}
just like in \eqref{pathprobS}--\eqref{corpo} but now with
probabilities
\[
q_\beta^e(\omega) \equiv  \mbox{Tr }[P_{\alpha_n}^n \Lambda_{n-1}
(\ldots P_{\alpha_1}^1 (\Lambda_1 (P_{\alpha_0}^0  \, \rho_0 \,
P_{\alpha_0}^0))P_{\alpha_1}^1\ldots)]
\]

Of course, if $\gamma$ in \eqref{ham} is zero, then $U_t = U^S_t
\otimes U^R$ factorizes and the treatments of Sections
\ref{mama}-\ref{genc} reduces to the adiabatic case. The work is a
difference of energies (instead of a
path-dependent quantity) and there is no heat ($Q=0$).\\

\subsubsection{Quasi-static regime} We imagine then that the
evolutions $U_t$ are slow enough so that the system plus
reservoirs relax into an
equilibrium state with respect to $H_t$.\\
We think about the case of Section \ref{genc}.  Always,
\begin{equation}
p_{\beta}(\omega,e)=q_{\beta}(\omega)q_{\beta}(e|\omega)
\end{equation}
but in the quasi-static regime we have
\begin{equation}
p_{\beta}(\alpha_t,E_t)=q_{\beta}(\alpha_t)q_{\beta}(E_t|\alpha_t)
\end{equation}
which suffices to see that $ \langle e^{-\beta (E_n-E_0)} \rangle
(\omega)$ depends only on $(\omega_0,\omega_n)$. Again, there is
no path dependence in the work $W$.

\subsubsection{No time-dependence} Suppose that \eqref{ham} does not contain a
parameter dependence and that the time-evolution is homogeneous
($U_t\equiv
U$).\\
Then of course $\Delta F=0$. For the effective Markovian dynamics
of Section \ref{mama}, one sees immediately that  $W(\omega)=0$
 for each $\omega$ in \eqref{wo}. In the adiabatic case of
  Section \ref{adi} as well:
   $V_{\alpha_0}^0 - V_{\alpha_n}^n=0$ with $q_\beta^S-$ probability one
when we ask that in \eqref{pathprobS} the projections
$P^t_{\alpha_t}$ and the unitary evolutions mutually commute. In
the case of Section \ref{genc}, we can use conservation of energy
$V^i_{\alpha_i}+E_i=$ constant, as in the first law \eqref{work},
when we ignore the (boundary) interaction term $H^I$ in the energy
balance. In that case we again get $W(\omega)=0$.\\

\noindent{\bf Acknowledgment:} We are much indebted to Karel
Neto\v cn\'y for many useful discussions.

\bibliographystyle{plain}

\end{document}